\documentclass{article}
\usepackage{graphicx}
\usepackage{amssymb}

\begin{document}

\title{\vspace{-2cm} Foam-like structure of the Universe}
\author{A. A. Kirillov \\
\emph{Institute for Applied Mathematics and Cybernetics, 10 Ulyanova str.,}\\
\emph{Nizhny Novgorod, 603005, Russia} \and D. Turaev \\
\emph{Ben-Gurion University of the Negev, P.O.B. 653, Beer-Sheva 84105,
Israel} }
\date{}
\maketitle

\begin{abstract}
On the quantum stage spacetime had the foam-like structure. When the
Universe cools, the foam structure tempers and does not disappear. We show
that effects caused by the foamed structure mimic very well the observed
Dark Matter phenomena. Moreover, we show that in a foamed space photons
undergo a chaotic scattering and together with every discrete source of
radiation we should observe a diffuse halo. We show that the distribution of
the diffuse halo of radiation around a point-like source repeats exactly the
distribution of dark matter around the same source, i.e. the DM halos are
sources of the diffuse radiation.
\end{abstract}

\pagebreak

\section{Introduction}

Wheeler pointed out that at Planck scales topology of spacetime undergoes
quantum fluctuations \cite{wheeler}. In the present Universe such
fluctuations carry virtual character and do not lead to detectable topology
changes. However, in the past, the Universe went through the quantum stage
when the temperature exceeded the Planckian value and the fluctuations were
strong enough to form a non-trivial topological structure of space. In other
words, on the very early, quantum stage Universe had to have a foam-like
structure. During the cosmological expansion, the Universe cools, quantum
gravity processes stop, and the topological structure of space freezes.
There is no obvious reason why the resulting topology has to be exactly that
of $R^{3}$ --- relics of the quantum stage foam might very well survive,
thus creating a certain distribution of wormholes in space. In the present
paper we show that the whole variety of the observed Dark Matter (DM)
phenomena admits a straightforward interpretation in terms of the foam-like
topological structure of space. Moreover, the specific properties of the
foam that are read from the observed DM distribution coincide with those
that are derived theoretically from very basic physical principles: we show
that the actual distribution of DM sources corresponds to the ground state
of the linear field theory on the foamed space.

An arbitrary non-trivial topology of space can be described as follows.
Given a Riemanian 3D manifold $\mathcal{M}$, we take a point $O$ in it and
issue geodesics from $O$ in every direction. Then points in $\mathcal{M}$
can be labeled by the distance from $O$ and by the direction of the
corresponding geodesic. In other words, for an observer at $O$ the space
will look as $R^3$ (endowed with a metric lifted from $\mathcal{M}$). Given
a point $P\in \mathcal{M}$, there may exist many homotopically
non-equivalent geodesics connecting $O$ and $P$. Thus, the point $P$ will
have many images in $R^3$. The observer might determine the topology of $%
\mathcal{M}$ by noticing that in the observed space $R^3$ there is a
fundamental domain $\mathcal{D}$ such that every radiation or gravity source
in $\mathcal{D}$ has a number of copies outside $\mathcal{D}$. The actual
manifold $\mathcal{M}$ is then obtained by identifying the copies. In this
way, we may describe the topology of space $\mathcal{M}$ by indicating for
each point $r\in R^3$ the set of its copies $E(r)$, i.e. the set of points
that are images of the same point in $\mathcal{M}$. Most of the time, we
will simply speak about the images of points in $R^3$, without referring to $%
\mathcal{M}$.

Note that an observer ignorant of the actual topological structure of $%
\mathcal{M}$ will greatly overestimate the density of matter (as all the
gravity sources outside the fundamental domain $\mathcal{D}$ are fictitious
--- each of them is just an image of some point in $\mathcal{D}$ seen from
another direction). However, one cannot immediately apply the above picture
to the explanation of DM effects: the Dark Matter emerges on galaxy scales
while we do not see multiple images of galaxies densely filling the sky. Our
idea that allows to link the observed DM effects with the topological
structure of space is that the fundamental domain may be of such distorted
shape that the direct recovery of the actual topology of space by detecting
images of sources could be impossible. Indeed, the non-trivial topology at
present is a remnant of quantum fluctuations at the very early Universe, and
the randomness built in the structure of the original quantum foam can
survive the cosmological expansion. Namely, at the quantum stage the state
of the Universe was described by a wave function defined on the space of
Riemanian 3D manifolds. Once quantum gravity processes stop, the further
evolution of the wave function was governed by the cosmological expansion
only. It is highly unlikely that the expansion could led to a complete
reduction of the wave function, i.e. to singling out one definite
topological structure of the Universe. In other words, if at the end of
quantum gravity era the Universe was not in a particular topological quantum
eigenstate, it is not in such a state now. One cannot, therefore, speak
about a definite topological structure of space, i.e. assign a definite set $%
E(r)$ of images to every point $r\in R^{3}$. A point $r^{\prime }\in R^{3}$
can be an image of $r$ with a certain probability only, hence instead of a
discrete set of images, a smooth halo of images of every single point
appears.

Even if we want to believe that a definite (classical) topological structure
has happened to emerge out of the quantum foam, the randomness of this
structure will persist: the wormholes which remained as the quantum foam
tempered will be randomly cast in space. Moreover, we recall that a typical
wormhole is obtained as follows: the interior of two remote spheres is
removed from $R^{3}$ and then the surfaces of the spheres are glued together%
\footnote{%
one can imagine a more general construction as well, where a pair of more
complicated two-dimensional surfaces replaces the spheres}. Such wormhole
works like a conjugated couple of convex (spherical) mirrors, therefore a
parallel beam of geodesics diverges after passing through the wormhole.
Thus, if we place spherical wormholes randomly in $R^{3}$, the flow of
geodesics that pass through a large number of the wormholes will have a
mixing property (like the flow of Sinai billiard, or of Lorenz gas). For a
point-like source for radiation or gravity, it means that some portion of
photons/gravitons will be scattered by the spherical wormholes, which will
create a specific smooth halo around every single source.

In any case, no matter what is the exact origin of the randomness of the
topological structure of space, one can take such random structure into
account by introducing a certain measure on the space of all Riemanian
3D-manifolds $\mathcal{M}$. The observed topological or metric properties of
space are then obtained by averaging over this measure. Thus, for example,
an individual manifold $\mathcal{M}$ is defined by specifying, for any point 
$r^{\prime}\in R^3$ the set $E(r^{\prime})$ of its images (the points in $%
R^3 $ that represent the same point of $\mathcal{M}$). Averaging over all
manifolds $\mathcal{M}$, gives a distribution $\bar K(r,r^{\prime})$ of the
images of $r^{\prime}$: 
\begin{equation}  \label{kxy}
\bar K(r,r^{\prime})=\delta(r-r^{\prime}) + \bar b(r,r^{\prime}),
\end{equation}
where the first term corresponds to the point $r^{\prime}$ itself, while $%
\bar b(r,r^{\prime})$ is a certain smooth distribution of additional images
of $r^{\prime}$; namely, in the neighborhood of a point $r$ of volume $d^3r$
there is (on average) $\bar b(r,r^{\prime})d^3r$ images of $r^{\prime}$.

It means that a single particle of matter at the point $r^{\prime }$ is
always accompanied by a smooth density $\bar{b}(r,r^{\prime })$ of exactly
the same matter. This halo does not necessarily emit enough light to be
identified, but it will always contribute to gravity. Thus, if the halo is
not seen, it is detected by an anomalous behavior of the gravitation
potential of the point-source. Such anomalous behavior is indeed universally
observed starting with the galaxy scales, and constitutes the DM phenomenon.
The existence of a quite rigid dependence between the density of luminous
matter (LM) and the density of DM is a well-known observational fact. This
fact allows us to interpret the DM phenomenon as an indication of the random
topological structure of space, with formula (\ref{kxy}) giving 
\begin{equation}
\rho _{DM}(r)=\int \bar{b}(r,r^{\prime })\rho _{LM}(r^{\prime
})d^{3}r^{\prime }.  \label{brdm}
\end{equation}%
In fact, the simple law 
\begin{equation}
\bar{b}(r,r^{\prime })\sim |r-r^{\prime }|^{-2}\;\;\;\mbox{ at }%
\;\;|r-r^{\prime }|\geq R_{0}  \label{law}
\end{equation}%
(where $R_{0}$ is the galaxy scale) provides quite accurate description of
all known DM effects. In particular, it allows to recover the whole variety
of observed galaxy rotation curves \cite{KT06}. It is also consistent with
the observed fractal structure of the distribution of matter on large scales 
\cite{K06} -\cite{Fra3}.

Note that relations (\ref{brdm}),(\ref{law}) give a good description for the
observed DM phenomena, independently of a theoretical interpretation \cite%
{K06,KT06}. We will, however, show that in our picture where $\bar
b(r,r^{\prime})$ is an averaged characteristic of the topological structure
of space, empirical law (\ref{law}) acquires a basic physical meaning.

It is also important that in our interpretation the DM halo is not actually
dark. The image $r$ of a point $r^{\prime }$ represents the same physical
point, just seen from another direction. Therefore, if the source of gravity
at $r^{\prime }$ is also a source of radiation, all its images in the halo
will be luminous too. However, the halo radiation has a diffuse character
and the brightness is very low (the halo radiates a reflected light, in a
sense). In observations, relating the halo radiation to a particular point
source could be a very difficult task\footnote{%
We note that we neglect here the red shift of light. In the actual Universe
every ghost image has its own red shift, depending on the value of the
traversed optical path, which puts an additional problem in relating the
halo radiation to the point source.}. In fact, the presence of a significant
diffuse component in cosmic radiation, unidentified with any particular
source, is well known \cite{Sar}. Usually, the observed diffuse halos in
galaxies are attributed to reflection from dust, and the general diffuse
component is assumed to originate from very fade and remote galaxies, but it
has never been related to DM halos. However, it was very convincingly
demonstrated in \cite{DMpaper} that the observed DM/LM ratio within the
intracluster gas clouds is much less than that for galaxies. This
observation gives a strong argument in support of our theory of DM effects:
while for small and bright sources (galaxies) the luminosity of the halo is
filtered out by the observer and the halo appears to be dark, for the
extended radiation sources (cluster size plasma clouds) the diffuse halo
radiation comes from the same region of space and is automatically accounted
in the total luminosity of the cloud.

Indeed, we show below that the intensity of sources of radiation
renormalizes according to the following law: 
\begin{equation}  \label{tempr}
I_{total}(r)=I_{source}(r)+I_{halo}(r),
\end{equation}
where 
\begin{equation}  \label{temp}
I_{halo}(r)=\int \bar b(r,r^{\prime}) I_{source}(r^{\prime}) d^3r^{\prime},
\end{equation}
with the same $\bar b(r,r^{\prime})$ as in (\ref{brdm}). Therefore, in our
picture, the luminosity of the DM is always proportional to its density. The
gravitating halos of discrete light sources in the sky only appear to be
dark, because of their diffuse character.

From the physical standpoint the foamed space is a porous system. It means
that the coordinate volume, which comes out from the extrapolation of our
local (solar) coordinate system, always exceeds the actual physical volume
(due to the presence of wormholes). The ratio $V_{coord}/V_{phys}=Q$ defines
the porosity coefficient of the foamed space. When we use the extrapolated
coordinates we always overestimate (by the use of the Gauss divergence
theorem) the actual intensity of a source of gravity or of an incoherent
radiation. In gravity, the effect displays itself as the presence of Dark
Matter. Hence, the porosity coefficient of the foamed space $Q$ can be
related to the ratio of Dark Matter density to the density of baryons in the
Universe, i.e. $Q=\Omega _{DM}/\Omega _{b}$. Analogously, the same relation
holds true for the ratio of two components of radiation (diffuse background
and discrete sources), i.e. $Q=\Omega _{diffuse}/\Omega _{discrete}$. The
relation 
\[
\Omega _{DM}/\Omega _{b}\approx \Omega _{diffuse}/\Omega _{discrete} 
\]%
is the basic indication of a geometrical (topological) nature of DM effects.

We point out that certain models of the spacetime foam have already been
considered in the literature (e.g., see Refs. \cite{BK07,F} and references
therein). However the primary interest was there focused on setting
observational bounds on the possible foam-like structure at extremely small
scales (i.e., at very high energies) $\gtrsim 10^{2}L_{pl}$ (where $L_{pl}$
is the Planck length), while DM phenomena suggest that the characteristic
scale of the spacetime foam $L$ (and respectively of wormholes) should be of
the galaxy scale, e.g., of the order of a few $Kpc$. The rigorous bounds
obtained indicate that at small scales spacetime is extremely smooth up to
the scales $\gtrsim 10^{2}L_{pl}$, that was to be expected\footnote{%
Indeed, at those scales topology fluctuations have only virtual character
and due to renormalizability of physical field theories they should not
directly contribute to observable (already renormalized) effects. Topology
fluctuations were strong enough only during the quantum stage of the
evolution of the Universe, while the possible subsequent inflationary phase
should considerably increase all characteristic scales of the foam. By other
words, the relic foam - like structure of space may survive only on very
large scales.}. The common feature of such models is that photons, in
addition to the chaotic scattering, undergo also modified dispersion
relations, as it happens in all Lorentz violating theories with preferred
frames (i.e., \textquotedblleft Aether-like situations\textquotedblright ,
e.g., see Refs. \cite{Aether}) which should lead to a modification of the
CMB spectrum acoustic peaks. The foam-like structure discussed in the
present paper surely violates the Lorentz invariance and also leads to some
modification of dispersion relations. However the Lorentz invariance and the
standard dispersion relations violate only at galaxy scales ($L\sim $ of a
few $Kpc$) which are unimaginably larger than any photon wave length $%
\lambda =c/\omega $ detected. We recall that in the Friedman Universe $%
\lambda ,L\sim a(t)$, where $a(t)$ is the scale factor and the ratio $%
\lambda /L\ll 1$ remains constant up to the quantum era. Therefore, such a
modification cannot directly influence the CMB spectrum (though it surely
influences via the DM effects discussed).

\section{Random Topology of Space}

In order to set a general frame for the study of a foamed space, let us
start with a toy example where the space is a cylinder of radius $R$. The
metric is the same as for the standard flat Friedman model 
\begin{equation}
ds^{2}=dt^{2}-a^{2}(t)(dx^{2}+dy^{2}+dz^{2}),  \label{fs}
\end{equation}%
but one of the coordinates, say $z$, is periodic ($z+2\pi R=z$). In what
follows, for the sake of simplicity we neglect the dependence of the scale
factor on time in (\ref{fs}), i.e. consider the Minkowsky space as the
coordinate space. Thus the actual values of the coordinate $z$ run through
the fundamental region $z\in \lbrack 0,2\pi R]$. Such space can be equally
viewed as a portion of the ordinary $R^{3}$ between two plane mirrors (at
the positions $z=0$ and $z=2\pi R$). An observer, who lives in such space,
may use the extrapolated reference system (i.e., $z\in (-\infty ,\infty )$),
however he/she easily notices that all physical fields are periodic in $z$.
Consider the Newton's potential $\phi $ for a point mass $M$. In this space
the exact expression can be easily found from the standard Newton's
potential by means of the image method. Indeed, the periodicity in $z$ means
that instead of a single point mass $M\delta (r-r^{\prime })$ at the point $%
r^{\prime }$ the observer will actually see an infinite series of images 
\begin{equation}
\delta (r-r^{\prime })\rightarrow K\left( r,r^{\prime }\right)
=\sum_{n=-\infty }^{\infty }\delta (z-z^{\prime }+2\pi Rn)\delta \left(
x-x^{\prime }\right) \delta \left( y-y^{\prime }\right) ,
\end{equation}%
and the Newton's potential for a point source at $r^{\prime }=0$ takes the
form 
\begin{equation}
\phi =-GM\sum_{n=-\infty }^{\infty }1/\sqrt{\rho ^{2}+\left( z+2\pi
Rn\right) ^{2}}.
\end{equation}%
On scales $r\ll R$ we may retain only one term with $n=0$ and obtain the
standard Newton's potential for a point mass $\phi \sim -GM/r$, while for
larger scales $r\gg R$ the compactification of one dimension will result in
the crossover of the potential to $\phi \sim \frac{GM}{R}\ln r$ (note that
this is indeed the shape of the potential that one reads from the observed
galaxy rotation curves).

The anomalous behavior of gravity indicates that DM effects show up at this
model on the scale of distances of order $R$. Indeed, let us consider a box
of the size $L$ and evaluate the total dynamical mass within the box 
\begin{equation}
M_{tot}\left( L\right) =M\int_{L^{3}}K\left( r,0\right) dV=M\left( 1+[\frac{L%
}{2\pi R}]\right).
\end{equation}
Thus, if the observer is ignorant about the real topological structure of
space he should conclude the presence of some extra (odd) matter. The reason
is obvious, when we expand the coordinate volume it covers the physical (or
fundamental) region of space many times and we respectively many times
account for the same source (i.e., images of the actual source). Thus the
increase of the total mass is fictitious. In the simplistic model under
consideration the volume of the fundamental (physical) region behaves as $%
V_{phys}=L^{3}$ for $L<R$ and $V_{phys}=2\pi RL^{2}$ for $L>R$. We note that
at large distances $L\gg R$ the parameter $Q\left( L\right) =M_{tot}(L)/M-1$
can be used to estimate the actual value of the physical volume: $%
V_{phys}(L)=L^{3}/Q\left( L\right) $, i.e., $Q$ is the ``porosity
coefficient'' of space at scales $L\sim R$.

The space discussed above is rather simple: for an extended source we will
see a countable set of its images without distortion. Therefore, one can
easily detect the fundamental region of space and avoid consideration of
fictitious sources. In the case of a general foamed topological structure
this is hardly possible. Nevertheless, whatever the topological structure of
the manifold is, we can apply the method of images: every topology can be
achieved by introducing a certain equivalence relation in $R^3$ and gluing
equivalent points together. Thus, a space of non-trivial topology is
completely defined by indicating for every point $r^{\prime}\in R^3$ the set 
$E(r^{\prime})=\{f_1(r^{\prime}),f_2(r^{\prime}),\dots\}$ of the points
equivalent to it. In other words, a point source at a point $r^{\prime}$ in
the fundamental region is accompanied by a countable set of images, or
``ghost'' sources: 
\begin{equation}
\delta (r-r^{\prime })\rightarrow K\left( r,r^{\prime }\right) =\delta
(r-r^{\prime })+\sum_{f_i(r^{\prime})\in E(r^{\prime})}\delta
(r-f_i(r^{\prime}))  \label{bf}
\end{equation}%
where $f_{i}\left( r^{\prime}\right) $ is the position of the $i$-th image
of the source.

For example, consider any source for radiation $J(r,t)$. Then according to (%
\ref{bf}) the electromagnetic potential $A\left(0,t\right)$ is described by
the retarded potentials 
\begin{equation}
A=\frac{1}{c}\int \frac{J_{t-|r|/c}}{|r|}dV+\frac{1}{c}\sum_{i}\int \frac{%
J_{t-|f_{i}(r)|/c}}{|f_{i}(r)|}dV_{i}.
\end{equation}
The first term of this formula corresponds to the standard, ``direct''
signal from the source, while the sum describes the multiple scattering on
the topological structure of space. A similar formula is obtained for the
gravitational field.

It is clear that all physical Green functions for all particles acquire the
same structure 
\begin{equation}  \label{greenb}
G_{total}(0,r) =G_0\left(0,r\right) +\sum_{f_i(r)\in
E(r)}G_i\left(0,f_{i}(r)\right).
\end{equation}
Formally, one can use the standard Green functions, while the scattering
will be described by the bias of sources 
\begin{equation}
J_{total}\left(r,t\right)=J\left( r,t\right) +\int b\left(
r,r^{\prime}\right) J\left( r^{\prime },t\right) d^3r^{\prime},  \label{bj}
\end{equation}%
where $b\left( r,r^{\prime }\right) =K\left( r,r^{\prime }\right) -\delta
(r-r^{\prime })$, i.e. we excluded the actual point source. In gravity the
second term in (\ref{bj}) corresponds to the DM contribution (e.g., see \cite%
{K06}). We note that in general the bias $b\left( r,r^{\prime }\right)$ is
an arbitrary function of both arguments, which means that the nontrivial
topological structure is capable of fitting an arbitrary distribution of
Dark Matter.

The function $K(r,r^{\prime })$ unambiguously defines the topological
structure of the physical space. However, for a general foamed structure of
space (a gas of wormholes) this function has a quite irregular character,
i.e. it is not directly observable. One has to introduce a measure on the
space of all 3D-manifolds and average the function $K$ over this measure.
The resulting function 
\[
\bar K(r,r^{\prime}) =\delta(r-r^{\prime})+\bar b(r,r^{\prime}) 
\]
gives the (average) density, at the point $r$, of the images of the point $%
r^{\prime}$.

Because of the averaging, the irregularities are smoothed out, hence the
bias function $\bar{b}(r,r^{\prime })$ is observable. Indeed, the averaging
of (\ref{greenb}) and (\ref{bj}) gives 
\begin{equation}
G_{total}(0,r^{\prime })=G\left( 0,r^{\prime }\right) +\int \bar{b}%
(r,r^{\prime })G\left( 0,r)\right) d^{3}r  \label{greena}
\end{equation}%
for Green functions, and 
\begin{equation}
\rho _{total}\left( r,t\right) =\rho \left( r,t\right) +\int \bar{b}\left(
r,r^{\prime }\right) \rho \left( r{\prime },t\right) d^{3}r^{\prime }
\label{bja}
\end{equation}%
for the density of matter. Therefore, when we can distinguish two components
in the observed picture of the distribution of, say, gravity sources:
discrete sources and a diffuse background, the discrete sources can be
identified with the first term in the right-hand side of (\ref{bja}), i.e.
with \textquotedblleft actually existing\textquotedblright\ sources, while
the diffuse halo can be identified with the second term, \textquotedblleft
the images\textquotedblright . Then, by comparing the observed distribution $%
\rho (r^{\prime })$ of actual (discrete) sources with the observed DM
distribution 
\begin{equation}
\rho _{halo}(r)=\int \bar{b}\left( r,r^{\prime }\right) \rho (r^{\prime
})d^{3}r^{\prime },  \label{halob}
\end{equation}%
one can extract an information about the structure of the bias $\bar{b}$. In
fact, the homogeneity of the Universe requires from $\bar{b}$ to be a
function of $(r-r^{\prime })$ only (which means that the form of DM halos
does not, in general, depend on the position in space). In this case, the
Fourier transform of (\ref{halob}) gives 
\begin{equation}
\rho _{halo}(k)=\bar{b}(k)\rho (k),  \label{halobf}
\end{equation}%
which defines $\bar{b}$ uniquely. As we show in the next Section, the bias $%
\bar{b}$ extracted from the DM observations in this way has both a very
simple form and a transparent theoretical meaning.

Note that being an averaged characteristics, the bias $\bar b$ does not
determine the topology of space completely. Along with the one-point
distribution $\bar K(r,r^{\prime})$, one can consider joint distributions of
images for several sources: 
\[
\bar K_n(r_1,\dots,r_n;\;r^{\prime}_1,\dots,r^{\prime}_n) 
\]
which is the averaged density of the images of the points $%
r_1^{\prime},\dots,r_n^{\prime}$ at the points $r_1,\dots,r_n$. Only when
all the functions $K_n$, $n=1,2,\dots$, are determined, one will have a full
description of the structure of the foamed physical space. However, the
one-point bias functions $\bar b(r,r^{\prime})$ carries the most important
information.

Thus, consider a source of radiation, constantly emitting light with the
frequency $\omega$, i.e. we have a density of the EM current $%
J(r^{\prime})e^{i\omega t}$ such that 
\begin{equation}  \label{srcj}
\langle J(r_1^{\prime}) J^*(r_2^{\prime})\rangle
=\delta(r_1^{\prime}-r_2^{\prime}) I_{source}(r_1^{\prime}),
\end{equation}
where $I_{source}(r)$ is the spatial distribution of the intensity of the
source. In order to take into account the effects of the non-trivial
topology of space, $J(r)$ should be modified according to (\ref{bj}), i.e. $%
J(r_1) J^*(r_2)$ transforms into 
\[
\int K\left(r_1,r^{\prime}_1\right)K\left(r_2,r^{\prime}_2\right)
J(r_1^{\prime})J^*(r_2^{\prime}) d^3r^{\prime}_1d^3r^{\prime}_2=\int
K(r_1,r^{\prime})K(r_2,r^{\prime}) I_{source}(r^{\prime})d^3r^{\prime}. 
\]
Averaging over different topologies gives 
\begin{equation}  \label{promezh}
\left(J(r_1) J^*(r_2)\right)_{total}= \int \bar
K_2\left(r_1,r_2;\;r^{\prime}, r^{\prime}\right) I_{source}(r^{\prime})
d^3r^{\prime},
\end{equation}
where $\bar K_2(r_1,r_2;\;r^{\prime}, r^{\prime})$ is, by definition, the
joint distribution of a pair of images of the point $r^{\prime}$.

The points $r_1$ and $r_2$ can be images of the same point $r^{\prime}$ if
and only if they are images of each other. Therefore, $\bar
K_2\left(r_1,r_2;\;r^{\prime}, r^{\prime}\right)$ is proportional to $\bar
K(r_1, r_2)=\delta(r_1 - r_2)+ \bar b(r_1, r_2)$; more precisely 
\begin{equation}  \label{k2p}
\bar K_2\left(r_1,r_2;\;r^{\prime}, r^{\prime}\right)= \delta(r_1-r_2)\bar
K(r_1,r^{\prime})+ \bar b(r_1, r_2) P(r_1,r_2,r^{\prime})
\end{equation}
where we denote as $P(r_1,r_2,r^{\prime})$ the density at the point $r_2$ of
the distribution of images of the point $r^{\prime}$ under the condition
that the point $r_1\neq r_2$ is an image of $r_2$.

As we see from (\ref{promezh}),(\ref{k2p}), while the phases of the source
current $J(r^{\prime })$ are delta-correlated (see (\ref{srcj})), there
appear long-range correlations in the density of the total current --- due
to the term proportional to $\bar{b}(r_{1},r_{2})$ in the kernel $\bar{K}%
_{2} $. However, the characteristic wave length in $\bar{b}(r_{1}-r_{2})$ is
of order of galaxy size, i.e. it is unimaginably larger than the wave length 
$c/\omega $ of the light emitted. Therefore, the contribution of the
coherent part of the total current to the radiation is completely
negligible: by (\ref{promezh}),(\ref{k2p}) we find 
\[
\left( J(r_{1})J^{\ast }(r_{2})\right) _{total}=\delta (r_{1}-r_{2})\int 
\bar{K}\left( r_{1},r^{\prime }\right) I_{source}(r^{\prime })d^{3}r^{\prime
}+\;\mbox{
long wave terms}, 
\]%
which gives the following formula for the total intensity of sources (actual
plus ghost ones) 
\begin{equation}
I_{total}(r)=\int \bar{K}\left( r,r^{\prime }\right) I_{source}(r^{\prime
})d^{3}r^{\prime }=I_{source}(r)+\int \bar{b}\left( r,r^{\prime }\right)
I_{source}(r^{\prime })d^{3}r^{\prime }.  \label{main}
\end{equation}%
Comparing with (\ref{halob}), we see that the distribution of a diffuse
radiation background associated to a luminous source coincides with the
distribution of dark matter in the halo of the same source.

Note that for a non-stationary remote source of radiation the picture is
more complicated. A momentary pulse at some point will create a spherical EM
wave emanating from the point --- and from its images. On the front of the
wave only a small number of images will give an essential contribution,
namely those which have comparable and shortest optical paths. This will
lead to an interference picture on the front. We note that due to wormholes
the signal from some images can reach an observer even earlier than the
basic signal. Only with time elapsed, as the larger and larger number of
images contribute, the interference picture disappears, and the diffuse
radiation background given by (\ref{main}) establishes.

In conclusion of this section, we recall that the observed homogeneity and
isotropy of space require from the topological bias $\bar{b}(r,r^{\prime })$
that defines both the DM distribution (\ref{halob}) and the distribution (%
\ref{main}) of the sources of diffuse radiation to be the function of the
distance $|r-r^{\prime }|$ only: $\bar{b}(r,r^{\prime })=\bar{b}%
(|r-r^{\prime }|)$. The integral 
\begin{equation}
Q\left( L\right) =4\pi \int_{0}^{L}R^{2}\overline{b}\left( R\right) dR
\label{q}
\end{equation}%
characterizes then the distortion of the coordinate volume or the porosity
of space (i.e., $1/Q$ gives the portion of the fundamental region or the
volume of the actual physical space in a coordinate ball of the radius $L$).
In general there can be both a situation where $Q(L)$ tends to a finite
limit as $L\rightarrow \infty $ and then $Q(\infty )$ defines the total
amount of DM ($Q=\Omega _{DM}/\Omega _{b}=\Omega _{diffuse}/\Omega
_{discrete}$), and the case where $Q$ is unbounded. The last case indicates
the presence of a certain dimension reduction of space at large distances
(e.g. when $Q(L)\sim L^{\alpha }$ the dimension of the physical space
reduces to $D=3-\alpha $ \cite{K03}).

\section{Topological bias: empirical and theoretical approach}

In this Section we derive a formula for the bias function $\bar
b(|r-r^{\prime}|)$ and show that it fits the observed picture of DM
distribution quite well. While in empirical considerations it is more
convenient to view $\bar b(R)$ as a bias of sources (which means exploring
the laws (\ref{halob}) and (\ref{main})), we achieve more theoretical
insight when choose an equivalent description of the random topological
structure of space by means of the bias of Green functions (see (\ref{greena}%
)). This means that instead of saying that each material point is
accompanied by an infinite set of images, we say that each source excites an
infinity of fields. Indeed, on a connected manifold of non-trivial topology
there is an infinite number of geodesics connecting any two points. So the
light emitted at a point $P$ arrives at a point $Q$ by an infinite number of
non-homotopic ways. We may associate a separate EM field with each homotopy
class: each of the fields propagates independently, but they sum up when
interact with matter. When we describe things in $R^3$ by means of the bias
functions, we thus associate a separate field to each term in the right-hand
side of (\ref{greenb}). These terms differ by positions of the images $%
f_i(r) $. In our picture, where the topology is random, there is no
preferred position for the $i$-th image, hence we have a system of an
infinite number of fields $\{A_i\}$ which is symmetric with respect to any
permutation of them (in other words, the fields are identical).

It is widely believed that the effects of quantum gravity should lead to a
cut-off at large wave numbers. The cut-off at $\Lambda$ means that the
photons with wave numbers $|k|>\Lambda$ are never excited. We say that the
field does not exist at such $k$. One can describe a cut-off of a more
general form, by introducing a characteristic function $\chi(k)$: at $%
\chi(k)=1$ the field with the wave number $k$ exists, while at $\chi(k)=0$
it does not. Because of the renormalizability of all physical field
theories, the question of the determining exact form of the cut-off of a
given field is of little importance. However, for the system of an infinite
number of identical fields $\{A_i\}$ the cut-off function acquires a meaning.

Indeed, let us define $N(k)=\sum_{i}\chi _{i}(k)$ where the sum is taken
over all the fields $A_{i}$. Thus, $N(k)$ is the number density of fields
which exist (i.e. which are not forbidden to create particles) at the given
wave number $k$. Here, the existence of the cut-off means that $N(k)$ can be
finite for all $k$. As the fields sum up when interacting with the matter,
the values of $N(k)$ greater than $1$ lead to a stronger interaction than in
the case of a single field. For example, consider a Newtonian potential%
\footnote{%
for the relativistic generalization see Sec. 2 in Ref. \cite{K06}.} 
\[
\Delta \phi =4\pi \gamma \rho . 
\]%
In the Fourier representation we have 
\begin{equation}
\phi (k)=\frac{-4\pi \gamma }{k^{2}}\rho (k).  \label{newp}
\end{equation}%
If there exist $N(k)$ identical Newtonian gravity fields with the wave
number $k$, each of them satisfies (\ref{newp}), while the effective
potential (that which acts on matter) is given by $\phi
_{eff}(k)=\sum_{i=1}^{N(k)}\phi _{i}(k)$ and satisfies, therefore, 
\[
\phi _{eff}(k)=\frac{-4\pi \gamma }{k^{2}}N(k)\rho (k). 
\]%
This is equivalent to a renormalization of the source density 
\[
\rho (k)\rightarrow N(k)\rho (k), 
\]%
and comparing with (\ref{halobf}) gives 
\[
N(k)-1=\bar{b}(k). 
\]%
Thus, the Fourier transform $\bar{b}(k)$ of the topological bias function
can be interpreted as the excessive number density of fields (gravity or EM)
at the wave number $k$, i.e. it is determined via a cut-off function.

Although the problem of determining the exact shape of the cut-off is
usually considered hopeless because the full quantum gravity theory has not
been developed, an approach developed in \cite{K99} allows one to derive
possible types of cut-off by means of simple thermodynamical models. For
example, assume that the energy density and the total excessive number
density of fields $\mathcal{N}=\int (N(k)-1)d^{3}k$ are finite. We also
assume that $\mathcal{N}$ is a conserved quantity (along with the energy).
Then the shape of the function $N(k)$ is determined uniquely by the
condition that the system of the identical free fields is in the
thermodynamical equilibrium (one should only choose the statistics for the
fields and fix the values of thermodynamical parameters). Indeed, the state
of the system with $N(k)$ identical free fields at the wave number $k$ is
determined by the numbers $n_{i}(k),i=1,\dots ,N(k)$ of the particles with
the wave number $k$ for each field. In the case of Fermi statistics for the
fields (that has nothing to do with the statistics for the particles which
remains Bose), there cannot be more than one field in the given state, i.e.
for every given $k$ all the numbers $n_{i}(k)$ should be different. The
energy density at the wave number $k$ equals to $\omega
_{k}\sum_{i=1}^{N(k)}n_{i}(k)$, where $\omega _{k}$ is the energy of a
single particle; as we deal here with massless fields, we take $\omega
_{k}=|k|$ (we put $h=c=1$). In what follows we assume Fermi statistics for
the fields (Bose statistics leads to a similar result \cite{K03,KT02},
however the computations in Fermi case are simpler). Then, the state of the
lowest possible energy (\textquotedblleft the ground state\textquotedblright
) corresponds to $\{n_{1}(k),\dots ,n_{N(k)}(k)\}=\{0,1,\dots ,N(k)-1\}$.
This gives us the energy $|k|N(k)(N(k)-1)/2$ at the wave number $k$. The
total energy density is thus given by $\int \frac{|k|}{2}N(k)(N(k)-1)d^{3}k$%
. The ground state corresponds to the minimum of the total energy density.
As the total excessive number density of fields $\mathcal{N}=\int
(N(k)-1)d^{3}k$ is assumed to be conserved, the problem of finding $N(k)$
reduces to minimizing $\int |k|N(k)(N(k)-1)d^{3}k$ under the constraint $%
\int (N(k)-1)d^{3}k=constant$. This gives us 
\[
N(k)=1+\left[ \frac{\mu }{|k|}\right] ,
\]%
where the \textquotedblleft chemical potential\textquotedblright\ $\mu $ is
fixed by the value of $\mathcal{N}$. For the bias function $\bar{b}$ this
gives 
\begin{equation}
\overline{b}\left( k\right) =\left\{ 
\begin{array}{ll}
\displaystyle\frac{\mu }{|k|} & \mbox{for }\ \ |k|<\mu , \\ 
0 & \mbox{for }\ \ |k|>\mu .%
\end{array}%
\right.   \label{b3}
\end{equation}

One can make different assumptions and, perhaps, arrive at different
formulas for the bias. However, this simplest bias function provides a very
good description of the observed distribution of DM. Indeed, in the
coordinate representation bias (\ref{b3}) takes the form 
\begin{eqnarray}
\overline{b}\left( \vec{r}\right) &=&\frac{1}{2\pi ^{2}}\int\limits_{0}^{\mu
}\left( \bar{b}\left( k\right) k^{3}\right) \frac{\sin \left( kr\right) }{kr}%
\frac{dk}{k}=  \label{b} \\
&=&\frac{\mu }{2\pi ^{2}r^{2}}\left( 1-\cos \left( \mu r\right) \right) . 
\nonumber
\end{eqnarray}%
As it was shown in \cite{KT06}, by choosing $\mu =\pi /\left( 2R_{0}\right) $
where $R_{0}$ is of order of a galaxy size (i.e. a few Kpc), bias (\ref{b})
applied to spiral galaxies produces the pseudo-isothermal DM halo\footnote{%
This result is valid for a single galaxy, while in the presence of a
distribution of galaxies the resulting halo acquires the Burket - type form $%
\rho =\rho _{0}R_{C}^{3}/(R_{C}^{2}+r^{2})(R_{\ast }+r)+\rho _{H}$, where $%
\rho _{H}$ is a homogeneous background formed by all galaxies and $R_{\ast }$
is the scale at which DM halo merges to the homogeneous DM background (e.g.,
see for discussions Ref. \cite{K06}).} $\rho =\rho
_{0}R_{C}^{2}/(R_{C}^{2}+r^{2})$, where $R_{C}$ is the core radius which has
the order of the optical disk radius $R_{C}\sim R_{opt}$. We note that this
is in a very good agreement with the observations (see \cite{PSS}). In fact,
by fitting one parameter $\mu $ in accordance to Tully-Fisher law \cite{TF},
relations (\ref{halob}) and (\ref{b}) quite accurately represent the whole
variety of the observed galaxy rotation curves \cite{PSS,KT06} (we recall
that bias (\ref{b}) is derived from thermodynamical considerations, so it is
quite natural to allow the chemical potential $\mu $ fluctuate in space;
exact mechanisms governing these fluctuations are described in \cite%
{K06,KT06}).

From (\ref{q}),(\ref{b}) one can find that starting with the galaxy scale
the porosity of space behaves as $Q\left(r\right) \sim r/R_{0}$. Thus the
total dynamical mass for a point source within the radius $r$ increases also
as 
\begin{equation}  \label{ling}
M\left( r\right) \sim M\left( 1+r/R_{0}\right).
\end{equation}
Importantly (see the previous Section), the same conclusion holds for the
luminosity of the point source (i.e., for a galaxy or an X-ray source).
Therefore, one can not immediately conclude from (\ref{ling}) a linear
growth of the ratio $M_{tot}(r)/M_b(r)$ of gravitational (dynamical or
lensing) to the barionic mass: the result depends on how much of diffuse
radiation is discarded at the observations.

Observations suggest that the number of baryons within the radius $r$
behaves as $N_{b}\left( r\right) \sim r^{D}$ with $D\simeq 2$ (see e.g.
Refs. \cite{Fra,Fra2,Fra3} where the $\simeq r^{2}$ behavior was reported up
to at least 200 Mpc). Thus, the observed baryonic density $\Omega _{b}$
falls inverse proportionally to the deepness of the observations and is well
below $1$. In the standard picture the total gravitational mass grows as $%
\sim R^{3}$, as it should be in a homogeneous Universe, so the linear growth
of $M_{tot}(r)/M_{b}(r)$ predicted by bias (\ref{b}) is indeed consistent
with observations. However, the linear growth starts to show up with the
scales larger than cluster size, while the reported mass to luminosity ratio
remains approximately the same on the galaxy scale and on the cluster scale.
To resolve the problem, we invoke the results of \cite{DMpaper} where there
was demonstrated that the intracluster gas clouds may not carry dark matter.
In our picture this is indeed the case, as the intracluster cloud is an
extended source of X-ray radiation, of size much larger than $R_{0}$. Thus,
the associated diffuse background sums up with the \textquotedblleft
direct\textquotedblright\ signal, so all the ghost sources of gravity that
lie within the cloud are visible as well. This means the absence of
\textquotedblleft dark\textquotedblright\ matter in the cloud or, in other
words, that the number of baryons in the cloud is greatly overestimated ---
most of the contribution to the cloud luminosity is given by the diffuse
halo, i.e. by fictitious sources due to the non-trivial topology of space.
It is easy to check that correcting the baryon density of the intracluster
gas in accordance with (\ref{main}), (\ref{b}) provides indeed the linear
growth of $M_{tot}(r)/M_{b}(r)$ starting right from the galaxy scale.

Note that at very large scales the diffuse radiation can hardly be separated
from the very faint sources. Therefore, the picture of the homogeneous
distribution of matter (i.e., of the Friedman Universe) is restored. In
fact, an arbitrary foam-like structure of space (i.e., any choice of the
bias $\overline{b}(r)$) agrees perfectly with the observational large-scale
homogeneity and isotropy of the Friedman Universe provided that the actual
physical volume $V_{phys}\left( r\right) =4/3\pi r^{3}/Q\left(r\right)$ (the
volume of the fundamental region of the coordinate space) is homogeneously
filled with matter. Indeed, in this case the number of actual sources within
the radius $r$ behaves as the physical volume $N_{b}\left( r\right) \sim
V_{phys}\left( r\right) \sim r^{3}/Q\left( r\right)$. Along with the actual
sources we always observe images (DM and diffuse radiation) and every source
produces $\Delta N\sim Q\left( r\right)$ additional images. Thus the total
number of images behaves always as $N_{b}\left( r\right) Q\left( r\right)
\sim r^{3}$, i.e., produces a homogeneous distribution.

\section{Conclusion}

In conclusion, we briefly repeat basic results. First of all the concept of
spacetime foam introduced by Wheeler can be crucial in explaining properties
of the present day Universe. The random (``foamed'') topological structure
leads to the fact that every discrete source in the sky should be surrounded
with a specific halo (a random distribution of images). We call this
phenomenon a topological bias of sources. In gravity such halo modifies the
standard Newton's law and appears as the Dark Matter phenomenon. In
particular, the Universal rotation curve (URC) constructed in \cite{KT06} on
the basis of the topological bias shows a very good fit to the empirical URC 
\cite{PSS}. We stress that in a general foamed space the bias $b\left(
r,r^{\prime }\right) $ is a random function of both arguments which means
that the form of the DM halo can arbitrary vary in space. By other words any
observed distribution of DM can be easily fitted by a proper choice of the
foamed structure. However, the simplest bias function which we derived
theoretically from a basic physical (thermodynamical) considerations seems
to give a quite accurate account of the DM effects in a huge range of
spatial scales.

As it was demonstrated in this paper, in the foamed space the halos around
discrete sources are actually not dark, but form the diffuse background of
radiation. Moreover, the ratio of the two components (the diffuse background
and discrete sources) is exactly the same as the ratio of DM and baryons ($%
\Omega _{DM}/\Omega _{b}=\Omega _{diffuse}/\Omega_{discrete}$).

We note that the foamed picture of our Universe allows to explain the
problem of missing baryons. Recall that the direct count of the number of
baryons gives a very small value $\Omega _{b}\sim 0.003$ for the whole
nearby Universe out to the radius $\sim 300h_{50}^{-1}Mpc$ e.g., see \cite%
{PS}. In our picture, this means only that at the radius $\sim
300h_{50}^{-1}Mpc$ the actual volume is ten times smaller, than in the
Friedman space ($V_{phys}\simeq 0.1V_{F}$), i.e. the actual density is ten
times bigger which reconciles the observed small baryon density with the
primordial nucleosynthesis constraints.

We stress that any homogeneously filled with matter foamed space (i.e., an
arbitrary choice of the bias function $b\left( r,r^{\prime }\right) $)
agrees perfectly with homogeneity and isotropy of the Universe and does not
contradict to the standard Friedman model. The general foamed Universe can
be viewed as the standard Friedman space filled with a gas of wormholes. In
such a picture the Large Scale Structure has an equilibrium character, for
it reflects the foamed topological structure of space (i.e., the
distribution of wormholes) formed during the quantum period of the evolution
of the Universe.

Finally, we have demonstrated that in a foamed space any non-stationary and
sufficiently remote signal is accompanied with a formation of a specific
interference picture at the front of the wave (stochastic interference)
which rapidly decays.

\section{Acknowledgment}

This research was supported in part by the joint Russian-Israeli grant
06-01-72023.

\end{document}